\documentclass[multphys,vecphys]{svmult}

\usepackage{makeidx}     % allows index generation
\usepackage{graphicx}    % standard LaTeX graphics tool
\usepackage{multicol}    % used for the two-column index
\makeindex             % used for the subject index
\begin{document}

\title*{Observations of Type Ia Supernovae, and Challenges for Cosmology}
\author{Weidong Li\inst{1}\and
Alexei V. Filippenko\inst{1}}
\institute{Department of Astronomy, University of California, Berkeley,
CA 94720-3411
\texttt{wli,alex@astro.berkeley.edu}
}
\maketitle

\begin{abstract}

Observations of Type Ia supernovae (SNe~Ia) reveal correlations between
their luminosities and light-curve shapes, and between their
spectral sequence and photometric sequence. Assuming SNe~Ia do not
evolve at different redshifts, the Hubble diagram of SNe~Ia may
indicate an accelerating Universe, the signature of a cosmological constant
or other forms of dark energy.  Several studies raise concerns about
the evolution of SNe~Ia (e.g., the peculiarity rate, the risetime,
and the color of SNe~Ia at different redshifts), but all these studies
suffer from the  difficulties of obtaining high-quality spectroscopy and
photometry for SNe~Ia at high redshifts.  There are also some troubling
cases of SNe~Ia that provide counterexamples to the observed correlations,
suggesting that a secondary parameter is necessary to describe the whole
SN~Ia family. Understanding SNe~Ia both observationally and theoretically
will be the key to boosting confidence in the SN~Ia cosmological results.

\end{abstract}

\section{Observations of SNe Ia and the Accelerating Universe}

Spectroscopic observations of nearby Type Ia supernovae (SNe~Ia) 
reveal that they can be
divided into several subclasses: the majority are the so-called
``normal" or ``Branch normal" SNe Ia (Branch, Fisher, \& Nugent 1993),
while the others are ``peculiar" SNe Ia which can be further divided into
SN 1991T-like or SN 1991bg-like objects (see Filippenko 1997, and references
therein). Li et al. (2001a) discuss SN 1999aa-like objects as another
potential subclass of the peculiar SNe~Ia. The classification is based on
the spectra of SNe Ia before or near maximum light: 
normal SNe~Ia show conspicuous features of Si~II, Ca~II, and other 
intermediate-mass elements (IMEs; e.g., S~II, O~I); SN 1991T-like objects show
unusually weak IME lines, yet prominent high-excitation features of 
Fe~III; SN 1991bg-like objects have strong IME features, plus a broad
Ti~II absorption trough around 4100~\AA\, and enhanced Si~II/Ti~II
$\lambda$5800 absorptions. SN 1999aa-like objects are similar to the
SN 1991T-like ones, but with significant Ca~II H \& K absorption lines.

Photometric observations of nearby SNe~Ia also reveal a correlation
between the peak luminosity and light-curve shape (LLCS correlation, hereafter).
This was first convincingly demonstrated by
Phillips (1993), and subsequently exploited by Hamuy et al.  (1996a), Riess,
Press, \& Kirshner (1996), Perlmutter et al. (1997), and Phillips et
al. (1999). The slower, broader light curves are intrinsically brighter at peak
than the faster, narrower light curves. Various parameters have been proposed
to quantify the ``speed" of the light curve, such as $\Delta m_{15}(B)$ (the decline
in magnitudes between peak brightness and 15 days later in the $B$ band), 
$\Delta$ (the difference in magnitudes between the peak brightness of a
SN~Ia and a nominal standard SN~Ia), $s$ (``stretch factor," the amount
of stretch applied to the light curve (generally $B$ band) of a SN~Ia to match 
those of a nominal standard SN~Ia), and various empirical methods have
been developed to calibrate the peak absolute magnitudes of SNe~Ia (Phillips
1993; Hamuy et al. 1996a; Riess, Press, \& Kirshner 1996; Perlmutter
et al. 1997; Jha 2002;  Wang et al. 2003). The multi-color light-curve shape 
(MLCS) method, for example, has demonstrated the ability to achieve a 
scatter in the calibrated absolute magnitudes of SNe~Ia to $\sim$0.15 mag
(Riess, Press, \& Kirshner 1996).

By assuming that the observed correlation for nearby SNe~Ia also applies to
the objects at high redshift, utilizing the empirical calibration methods 
developed from nearby SNe~Ia, and studying the Hubble diagram for SNe~Ia
at both low and high redshifts, the High-z SN Search Team (Schmidt
et al. 1998) and the SN Cosmology Project (Perlmutter et al. 1997) 
have measured that high-redshift SNe~Ia are fainter 
than expected, and interpreted this result as evidence that the 
expansion of the Universe is accelerating, due perhaps to a non-zero 
cosmological constant or some other forms of dark energy (e.g., Riess et al.
1998; Perlmutter et al. 1999; Tonry et al. 2003; Knop et 
al. 2003). 

\section{Challenges of the Accelerating Universe}

The observational fact is that the SNe~Ia at $z \approx 0.5$ are $\sim$0.28 mag 
dimmer ($\sim$14\% farther) than expected in a Universe with $\Omega_M = 0.3$ and no 
cosmological constant. Besides the interpretation of a positive cosmological
constant, other possible alternatives have been proposed as follows.

\begin{enumerate}

\item{Luminosity evolution. Two important questions are (a) whether 
the high-redshift SNe~Ia follow the same LLCS
correlation for the nearby objects, and (b) whether the SNe~Ia at $z \approx 0.5$ 
are intrinsically fainter than nearby SNe~Ia by 0.28 mag after correction for the 
LLCS correlation.} 

\item{Interstellar dust, which produces more extinction of the high-redshift 
SNe~Ia and makes them look apparently fainter.}

\item{Selection bias: preferentially fainter SNe~Ia are observed
at high redshift.}

\item{Gravitational lensing: the inhomogeneous distribution of matter
in the Universe deamplifies the observed brightness of most high-redshift 
SNe~Ia.}

\end{enumerate}

Among these, ordinary dust is not considered a viable option, as it introduces
too much reddening in the colors of SNe~Ia at high redshift (Riess
et al. 2000) and more dispersion in the distance measurements than is currently
observed. ``Grey" dust (Aguirre 1999a,b), which leaves little or no imprint
on the spectral energy distribution of a SN~Ia, could be more pernicious, but
the amount of grey dust required to explain the faintness of high-redshift
SNe~Ia would also distort the cosmic microwave background, an effect 
which has not been seen. Moreover, it still introduces more distance
dispersion than is currently observed. Tonry et al. (2003)
also provided evidence that a systematic effect which goes
as a power law of $(1+z)$, such as extinction by dust, is not likely to match
the SN~Ia data.  No known selection bias favors detection of
intrinsically fainter SNe~Ia at high redshift. Malmquist bias may afflict
the nearby SN~Ia sample so 
preferentially brighter objects were observed, but
the effect is shown to be small (Riess et al. 1998; Perlmutter et al. 1999).
Moreover, the high-redshift SN~Ia sample should be subject to the same bias 
to a greater extent, since most high-redshift SN searches are
magnitude-limited. 
Gravitational lensing deamplification (Metcalf 1999; Barber 2000), typically
$\sim$ 2\% at $z = 0.5$, is much smaller than the cosmological effect. 

Luminosity evolution is arguably the most serious challenge to the cosmological
interpretation of high-redshift SNe~Ia.  The cause of this is somewhat 
embarrassing: despite being the most luminous type of SN, SNe~Ia have not been
completely understood theoretically (see Leibundgut 2001 for a review), or
have the progenitor system conclusively identified (see Livio 2000 for a
review). Consequently, theory cannot provide conclusive guidance on
whether or how SNe~Ia and their progenitor systems evolve at different redshifts. 
Nevertheless, theorists have provided some insights into this question 
by studying the effects of metallicity and the C/O ratio of a white dwarf (WD),
the agreed precursor to a SN~Ia (e.g., von Hippel, Bothun, \& Schommer 1997;
H\"oflich, Wheeler, \& Thielemann 1998;
Umeda et al. 1999; Nomoto et al. 2003). H\"oflich, Wheeler, \& Thielemann (1998)
suggested that the effect of changing metallicity on the rest-frame visual and 
blue light curves is small, and as the C/O ratio of a WD becomes
progressively lower at higher redshift, the luminosity of the resulting SN~Ia 
becomes brighter for the same light-curve shape, an effect that is contrary
to the cosmological result from SNe~Ia. Umeda et al. (1999) and Nomoto et al.
(2003) use the variation in the C/O ratio in the WD to explain the distribution
of SN~Ia brightness, but suggest that the diversity can be normalized by
applying the LLCS correlation.

Answers to the question of whether SNe~Ia evolve have also been sought from 
observations of them at different redshifts. The nearby sample is an
excellent laboratory for studying possible luminosity evolution, since SNe~Ia
have been observed in a wide range of host-galaxy morphologies including
ellipticals, spirals, irregulars, and dwarf galaxies. In fact, the range of
metallicity, stellar age, and interstellar
environments probed by the nearby SN~Ia sample
is much greater than the mean evolution in these properties for individual
galaxies between $z = 0$ and $z = 0.5$. Some variation of the observed
characteristics of SNe~Ia has been noticed; for example, luminous events occur 
preferentially in metal-poor environments (Hamuy et al. 2000), and the 
luminosity of SNe~Ia correlates with the projected distance from the host
nucleus (Wang, H\"oflich, \& Wheeler 1997). However, after correction for the LLCS
correlation and extinction, the observed residuals 
from the Hubble flow do not correlate with host-galaxy morphology 
or the projected radial distances (Riess 2000; Sullivan et al. 2003).
This suggests that the LLCS correlation
applies to a wide range of stellar environments and is a strong argument 
against significant evolution to $z = 0.5$ (Schmidt et al. 1998).

The empirical test of luminosity evolution at high redshift has been
focused on getting high-quality spectra and light curves, and comparing
them with those of nearby SNe~Ia. The assumption of this test is that 
significant luminosity evolution would be accompanied by other visibly
altered observables of the SNe. Comparison of high-quality spectra between
nearby and high-redshift SN~Ia have 
revealed remarkable similarity (Riess et al. 1998; Perlmutter et al. 1999;
Coil et al. 2000; Tonry et al. 2003). Riess et al. (2000) also
obtained the rest-frame $I$-band light curve of the high-redshift SN~Ia 1999Q,
which displayed the secondary maximum that is typical of normal nearby
SNe~Ia.

To date, there is no clear, direct evidence that suggests 
significant luminosity evolution for SNe~Ia at different redshift.
To rest the case of luminosity evolution, however, we need to fully
understand the models of SNe~Ia and how they evolve at different
redshifts. Without a firm 
theoretical footing, we must conservatively demand that 
{\bf all} observables of high-redshift SNe~Ia be statistically consistent
with their nearby counterparts. 

In the following sections, we discuss in more detail some of the recent
comparisons
done on the observables and characteristics of SNe~Ia at different redshifts,
which mag suggest possible differences between high-redshift SNe~Ia and their
nearby counterparts. 

\subsection{Peculiarity Rate at Different Redshifts}

The rate of ``peculiar SNe Ia" in the nearby sample has been recently measured
by Li et al.  (2001a).  They used a {\bf distance-limited} sample of SNe~Ia from
the Lick Observatory Supernova Search (LOSS; Filippenko et al. 2001) 
and performed Monte Carlo simulations
(Li, Filippenko, \& Riess 2001) demonstrating that essentially
all SNe~Ia should have been discovered in
the sample galaxies of LOSS. Within this unbiased sample they found a rate
of $\sim$20\%  for the
SN 1991T/1999aa-like objects and $\sim$16\% for the SN 1991bg-like objects, for
a total peculiarity rate of $\sim$36\%. 

However, in the now more than 100
spectroscopically classified SNe~Ia at high redshift, there has
not been a single unambiguously peculiar SN~Ia reported. While the lack of SN 1991bg-like 
objects could be explained by their intrinsic faintness and low expected
rate in {\bf magnitude-limited} searches, the lack of SN 1991T/1999aa-like objects is puzzling. 

Li et al. (2001a) offered several possible explanations for the difference
between the peculiarity rate of SNe~Ia at different redshifts: 
extinction toward the SN 1991T/1999aa-like objects, difficulty in
identifying peculiarities in poor-quality spectra of the high-redshift
SNe~Ia, and most importantly, the ``age bias": the peculiarity of
SN 1991T/1999aa-like objects can only be easily identified in early-time 
spectra.  This same bias may also
explain why the Cal\'an/Tololo survey, a nearby magnitude-limited SN search,
yielded no SN 1991T/1999aa-like objects among 29 SNe~Ia (Hamuy et al. 1996b).

If, however, these observational biases are not to blame, the absence of
peculiar SNe~Ia at high redshift could result from an evolution of the 
population of progenitor systems: certain progenitor channels at high
redshift may be lost due to a redshift-dependent variation in 
the mass, composition, and metallicity of SN~Ia progenitors (e.g., 
Ruiz-Lapuente \& Canal 1998; Livio 2000). It is thus an important challenge to
observationally identify some peculiar SNe~Ia at high redshifts, 
to definitively rule out luminosity evolution as the cause of the difference
in the peculiarity rate of SNe~Ia at different redshifts. 

\subsection{Risetime at Different Redshifts}

The risetime is defined as the time interval between the explosion and the
maximum brightness of a SN~Ia. Precise knowledge of the SN~Ia risetime, which
is sensitive to the ejecta opacity and the distribution of $^{56}$Ni, provides
constraints on models of SN~Ia progenitors.  A comparison of the
risetime for high-redshift and nearby SNe~Ia is thus a valuable test of
luminosity evolution. 

The risetime for the nearby SNe~Ia was measured by Riess et al. (1999).
They collected about 25 measurements of SNe~Ia between 10 and 18 days before
$B$ maximum, normalized them to a fiducial risetime curve, 
and measured a risetime of $19.98\pm0.15$ days. A preliminary risetime for the 
high-redshift SNe Ia was measured by Goldhaber (1998) from the Supernova
Cosmological Project (SCP) data as $17.50\pm0.40$ days, which is discrepant 
from the nearby risetime at a statistical likelihood greater than 
99.99\% (5.8$\sigma$). Aldering, Knop, \& Nugent (2000), however, refined
the risetime for the high-redshift SNe~Ia in the SCP data
to $17.50\pm1.20$ days, a $\sim 2\sigma$ difference from the nearby measurement.
They also suggested that under extreme situations the risetime could be biased
up to 2--3 days due to observational biases and fitting methods. 

It remains to be seen whether the risetimes of high-redshift and nearby
SNe~Ia are statistically inconsistent when a better risetime measurement
is derived for the high-redshift SNe~Ia.
It should also be noted that even if the two risetimes are
inconsistent with each other, it is unclear whether the difference in risetime
could be translated into a difference in peak luminosity: most current theoretical models 
have difficulties in reproducing the observed risetimes and the correlation
between risetime and peak luminosity. 

\subsection{Intrinsic Color at Different Redshifts}

To date, there is no consensus on the precise intrinsic colors of SNe~Ia with
different photometric behaviors. As a result, there is not a good theoretical
or empirical method to accurately determine host-galaxy extinction to SNe~Ia,
and observers often have to resort to priors such as applying the
Galactic reddening law to the host galaxies of SNe~Ia, and assuming all 
SNe~Ia have the same intrinsic color at maximum. Phillips et al. (1999) proposed
a method to estimate the host-galaxy reddening to a SN~Ia by using its color
at 30--60 days past maximum, but Li et al. (2001b) showed that this method
does not apply to all objects.  The extinction
correction is a major source of uncertainty in the current empirical fitting
methods. To circumvent this difficulty, people often use subsamples that
are likely to have low extinction at both low and high redshifts.
Fortunately, Hatano, Branch, \& Deaton (1998) showed that most SNe~Ia
should have low extinction. 

Different methods for treating the extinction correction in the fitting process
yield different results in the intrinsic color comparison of SNe~Ia at a range of
redshifts. Leibundgut (2001) and Falco et al. (1999) suggested that there
is evidence from the $E(B-V)$ values in Riess et al. (1998) that high-redshift 
SNe Ia are statistically bluer than their nearby counterparts, but
analysis by Perlmutter et al. (1999) showed no such effect, nor did a recent
compilation of 11 high-redshift SNe Ia observed by the {\it Hubble Space
Telescope} (Knop et al. 2003). 
The problem of extinction correction will continue to plague the empirical
fitting methods until a better understanding of the intrinsic colors of SNe~Ia 
is achieved, and the influence of extinction corrections on the cosmological 
conclusions needs to be investigated in more detail. Drell, Loredo, 
\& Wasserman (2000),
for example, attributed the difference in colors to luminosity evolution,
while Knop et al. (2003) suggested that reasonable changes in colors do not
have a significant impact on the cosmological results. 

\subsection{Peculiar Nearby SNe Ia}

A fundamental assumption of the current empirical fitting methods is that
the light curves of all normal SNe~Ia can be represented by a single parameter
such as $\Delta m_{15}(B)$, $\Delta$, or $s$. However, there is
growing evidence that not all SNe~Ia form a one-parameter family. Branch (1987)
showed that normal SNe~Ia could have very different expansion velocities. 
Hamuy et al. (1996c) showed that some light curves with similar decline rates have
significant differences in particular details. Hatano et al. (2000) also
demonstrated that the spectroscopic diversity among SNe~Ia is multi-dimensional.

The SN 1991T-like and SN 1991bg-like objects, though categorized as peculiar
SNe Ia, generally follow the LLCS correlation and the spectrum--luminosity
sequence --- i.e., they seem to be an extension of the one-parameter
description of normal SNe~Ia. Some more disturbing cases of peculiar
nearby SNe~Ia that fail the one-parameter description are SN 2000cx
(Li et al. 2001; Candia et al. 2003), SN 2002cx (Li et al. 2003),
SN 2001ay (Nugent et al. in preparation), and SN 2002ic (Hamuy et al. 2003). 

The peculiarity of SN 2000cx is that its light curves cannot be fit well by
the existing fitting methods. There is an apparent asymmetry
in the $B$-band peak, in which the premaximum brightening is relatively fast
(similar to that of the normal SN 1994D), but the postmaximum decline is
relatively slow (similar to that of the overluminous SN 1991T). SN 2000cx
has very blue colors and also unique spectral evolution. Its premaximum
spectra are similar to those of SN 1991T-like objects, but the high-excitation
Fe~III lines remain prominent until well after maximum. The expansion velocities
derived from the absorption features are unusually high and evolve
differently than normal. Though it has a slow
light curve, its estimated luminosity is average (Li et al. 2001)
or even slightly subluminous (Candia et al. 2003). 

SN 2002cx has many properties that are the opposite of those of SN 2000cx.
It has a premaximum spectrum similar to that of SN 1991T, a decline rate in
the $B$-band similar to that of normal SNe~Ia, but a luminosity 
similar to that of the very subluminous SN 1991bg. It has a very red color
evolution, and has extremely low expansion velocities measured from spectral
features. The $R$ and $I$-band light curves have a peculiar plateau phase 
around maximum. The late-time decline rate in all $BVRI$ bands is unusually
slow. SN 2003gq (Filippenko \& Foley 2003) may be another event that is 
similar to SN 2002cx.

SN 2001ay has a normal near-maximum spectrum, except that it has very
high expansion velocities. The light curves of SN 2001ay are the slowest ever
recorded, yet it has a normal luminosity.  It also has a peculiar red color
evolution until 30 days after maximum. 

SN 2002ic is the only SN~Ia to have shown direct evidence of SN ejecta
interacting with the circumstellar medium (CSM).
Its near-maximum spectrum is similar to that of SN 1991T, but diluted in
strength. There are remarkable Balmer lines in later spectra,
with $H\alpha$ showing an unresolved component (FWHM $<$ 300 km s$^{-1}$) 
superimposed on a broad resolved base (FWHM $\approx$ 1800 km s$^{-1}$), similar to
those observed in Type IIn SNe (Filippenko 1997). The 
spectral features and photometric behavior of SN 2002ic suggest that it 
has a very dense CSM. Based on these observations, Hamuy et al. (2003) ruled out
the double-degenerate model for SN 2002ic, and suggested that the progenitor system
involves a white dwarf and an asymptotic giant branch star. 
Livio \& Riess (2003), however, argued that the opposite may be true:
SN 2002ic results from a rare circumstance in which the SN~Ia ejecta interact
with the previously ejected common envelope of a double-degenerate system. 

Although the frequency of these peculiar SNe~Ia is low, and statistically
they will not challenge the established empirical correlations, we need to
understand why they are peculiar, and how they can be fit into
the whole picture of SN~Ia theories and observations. These objects
with unusual properties might represent the general models of 
SNe~Ia under extreme conditions, and studying them will provide clues to
the theoretical models and progenitor systems. It is interesting to note
that three of the four peculiar SNe~Ia (SNe 2000cx, 2002cx, and 2001ay) all
have very unusual expansion velocities, and three (SNe 2000cx, 2002cx, and
2002ic) show spectral features similar to those of SN 1991T. The subclass of 
SN 1991T/1999aa-like objects may thus be more heterogeneous than other SNe~Ia,
and objects with unusual expansion velocities should be treated with caution
when used as cosmological tools.

\section{Conclusions}

Many alternatives have been proposed to explain the SN~Ia data at different
redshifts, but so far none has seriously challenged the accelerating Universe
result.  We have found no clear, direct evidence
that SNe~Ia at different redshifts evolve, though some studies show that there
may exist some differences in their peculiarity rate, risetime, or colors. 

The key to boosting confidence in the cosmological results from SNe~Ia
is to understand SNe~Ia both theoretically and observationally.
We need to theoretically identify the elusive progenitor systems
for SNe Ia, and find out the cause of the diversity of SNe~Ia. 
Similarly, we need to continue to search for SN/CSM interactions such as
that observed in SN 2002ic, and place stringent constraints on the accretion
history of SN~Ia progenitors. We also need to re-examine existing observations
of SNe~IIn, to investigate whether SN 2002ic is an isolated case,
or whether some additional SNe~IIn are actually SNe~Ia with strong SN/CSM interaction.
For nearby SNe~Ia, we need to develop better methods to measure host-galaxy 
extinction than currently available, study the 
environmental effects, find more empirical correlations, and develop a
subclassification scheme that possibly links to different progenitor
channels. We should continue to study those SNe~Ia that are clearly
discrepant. For high-redshift SNe~Ia, we need to identify some peculiar
SN 1991T-like or SN 1991bg-like objects, get 
better risetime measurements, obtain more high-quality spectra
and light curves, and compare them with those of nearby SNe~Ia. 
The ESSENCE project (e.g., Garnavich et al. 2002), SNAP satellite
(http://snap.lbl.gov/), and the
higher-z project (Riess 2002) are prime
examples of current and future extensive studies of high-redshift SNe~Ia.

\acknowledgement

Our research is currently supported by NSF grants AST-0206329 and AST-0307894,
by the Sylvia \& Jim Katzman Foundation, and
by NASA grants GO-8641, GO-9114, and GO-9352 from the Space Telescope
Science Institute, which is operated by AURA, Inc., under NASA contract
NAS 5-26555. We thank the conference organizers for partial travel funds.

\printindex

\begin{thebibliography}{99.}

\bibitem{}Aguirre, A. N.: Astrophys. Journ.,\textbf{512}, L19 (1999a)
\bibitem{}Aguirre, A. N.: Astrophys. Journ.,\textbf{525}, 583 (1999b)
\bibitem{}Aldering, G., Knop, R., \& Nugent, P.: Astron. Journ. \textbf{119}, 2110 (2000)
\bibitem{}Barber, A. J.: Mon. Not. Roy. Astron. Soc. \textbf{318}, 195 (2000)
\bibitem{}Branch, D.: Astrophys. Journ. \textbf{316}, L81 (1987)
\bibitem{}Branch, D., Fisher, A., \& Nugent, P.: Astron. Journ. \textbf{106}, 2383 (1993)
\bibitem{}Candia, P., et al.: Pub. Astron. Soc. Pac. \textbf{115}, 277 (2003)
\bibitem{}Coil, A., et al.: Astrophys. Journ. \textbf{544}, L111 (2000)
\bibitem{}Drell, P. S., Loredo, T. J., \& Wasserman, I.: Astrophys. Journ. \textbf{530}, 593 (2000)
\bibitem{}Falco, E., et al.: Astrophys. Journ. \textbf{523}, 617 (1999)
\bibitem{}Filippenko, A. V.: Ann. Rev. Astron. Astrophys. \textbf{35}, 309 (1997)
\bibitem{}Filippenko, A. V., \& Foley, R.: IAU Circ. 8211 (2003)
\bibitem{}Filippenko, A. V., Li, W. D., Treffers, R. R., \& Modjaz, M.: in
   {\bf Small-Telescope Astronomy on Global Scales}, eds.
   W. P. Chen, C. Lemme, \& B. Paczy{\' n}ski (San Francisco: Astron. Soc. Pac.), 121 (2001)
\bibitem{}Garnavich, P. M., et al.: B.A.A.S. \textbf{201}, 7809 (2002)
\bibitem{}Goldhaber, G.: B.A.A.S. \textbf{193}, 4713 (1998)
\bibitem{}Hamuy, M., et al.: Astron. Journ. \textbf{112}, 2391 (1996a)
\bibitem{}Hamuy, M., et al.: Astron. Journ. \textbf{112}, 2408 (1996b)
\bibitem{}Hamuy, M., et al.: Astron. Journ. \textbf{112}, 2438 (1996c)
\bibitem{}Hamuy, M., et al.: Astron. Journ. \textbf{120}, 1479 (2000)
\bibitem{}Hamuy, M., et al.: Nature \textbf{424}, 651 (2003)
\bibitem{}Hatano, K., Branch, D., \& Deaton, J.: Astrophys. Journ. \textbf{502}, 177 (1998)
\bibitem{}H\"oflich, P., Wheeler, J. C., \& Thielemann, F.-K.: Astrophys. Journ. \textbf{502}, 177 (1998)
\bibitem{}Jha, S.: PhD thesis, Harvard Univ. (2002)
\bibitem{}Knop, R. A., et al.: Astrophys. Journ., in press (astroph/0309368) (2003)
\bibitem{}Leibundgut, B.: Ann. Rev. Astron. Astrophys. \textbf{39}, 67 (2001)
\bibitem{}Li, W., Filippenko, A. V., \& Riess, A. G.: Astrophys. Journ.
\textbf{546}, 719 (2001)
\bibitem{}Li, W., et al.: Astrophys. Journ. \textbf{546}, 734  (2001a)
\bibitem{}Li, W., et al.: Pub. Astron. Soc. Pac. \textbf{113}, 1178 (2001b) 
\bibitem{}Li, W., et al.: Pub. Astron. Soc. Pac. \textbf{115}, 453 (2003)
\bibitem{}Livio, M.: in {\bf Type Ia Supernovae, Theory and Cosmology},
eds. J. C. Niemeyer \& J. W. Truran (Cambridge University Press), 33 (2000)
\bibitem{}Livio, M., \& Riess, A. G.: Astrophys. Journ. \textbf{594}, L93 (2003)
\bibitem{}Metcalf, R. B.: Mon. Not. Roy. Astron. Soc. \textbf{305}, 746 (1999)
\bibitem{}Nomoto, K., et al.: in {\bf From Twilight to Highlight: The 
Physics of Supernovae}, eds. W. Hillebrandt \& B. Leibundgut (Berlin: Springer), 115 (2003)
\bibitem{}Perlmutter, S., et al.: Astrophys. Journ. \textbf{483}, 565 (1997)
\bibitem{}Perlmutter, S., et al.: Astrophys. Journ. \textbf{517}, 565 (1999)
\bibitem{}Phillips, M. M.: Astrophys. Journ. \textbf{413}, L105 (1993)
\bibitem{}Phillips, M. M., et al.: Astron. Journ. \textbf{118}, 1766 (1999)
\bibitem{}Riess, A. G., Press, W. H., \& Kirshner, R. P.: Astrophys. Journ. \textbf{473}, 88  (1996)
\bibitem{}Riess, A. G., et al.: Astron. Journ. \textbf{116}, 1009 (1998)
\bibitem{}Riess, A. G., et al.: Astron. Journ. \textbf{118}, 2675 (1999)
\bibitem{}Riess, A. G.: Pub. Astron. Soc. Pac. \textbf{112}, 1284 (2000)
\bibitem{}Riess, A. G., et al.: Astrophys. Journ. \textbf{536}, 62 (2000)
\bibitem{}Riess, A. G.:  B.A.A.S. \textbf{201}, 3901 (2002)
\bibitem{}Ruiz-Lapuente, P., \& Canal, R.: Astrophys. Journ. \textbf{497}, 57 (1998)
\bibitem{}Schmidt, B. P., et al.: Astrophys. Journ. \textbf{507}, 46 (1998)
\bibitem{}Sullivan, M., et al.: Mon. Not. Roy. Astron. Soc. \textbf{340}, 1057 (2003)
\bibitem{}Tonry, J. L., et al.: Astrophys. Journ. \textbf{594}, 1 (2003)
\bibitem{}Umeda, H., et al.: Astrophys. Journ. \textbf{522}, 43 (1999)
\bibitem{}von Hippel, T., Bothun, G. D., \& Schommer, R. A.: Astron. Journ. \textbf{114}, 1154 (1997)
\bibitem{}Wang, L., H\"oflich, P., \& Wheeler, J. C.: Astrophys. Journ. \textbf{483}, 29 (1997)
\bibitem{}Wang, L., Goldhaber, G., Aldering, G., \& Perlmutter, S.: Astrophys. Journ. \textbf{590}, 944 (2003)

\end{thebibliography}
\end{document}